\documentclass[aps,prl,reprint,groupedaddress,showpacs]{revtex4-1}

\usepackage{fancyhdr}
\usepackage{graphicx}
\usepackage{amsmath,amsthm,amssymb}
\usepackage{float}
\usepackage[hidelinks=true]{hyperref}
\usepackage{subcaption}
\usepackage{color,soul} 

\usepackage[compact]{titlesec}
\titlespacing{\section}{0pt}{1.0\baselineskip}{1.0\baselineskip}
\titlespacing{\subsection}{0pt}{0.8\baselineskip}{0.8\baselineskip}

\expandafter\def\expandafter\normalsize\expandafter{%
    \normalsize
    \setlength\abovedisplayskip{6pt}
    \setlength\belowdisplayskip{6pt}
}

\newcommand{\bd}[1]{\textbf{#1}}
\renewcommand{\vec}[1]{\boldsymbol{#1}}

\begin{document}


\title{Towards a `Thermodynamics' of Active Matter}


\author{S.C. Takatori}

\author{J.F. Brady}
\email[]{jfbrady@caltech.edu}

\affiliation{Division of Chemistry and Chemical Engineering, California Institute of Technology, Pasadena, California 91125, USA }



\date{November 14, 2014}

\begin{abstract}
Self-propulsion allows living systems to display unusual collective behavior.  Unlike passive systems in thermal equilibrium, active matter systems are not constrained by conventional thermodynamic laws.  A question arises however as to what extent, if any, can concepts from classical thermodynamics  be applied to nonequilibrium systems like active matter.  Here we use the new swim pressure perspective to develop a simple theory for predicting phase separation in active matter.  Using purely mechanical arguments we generate a phase diagram with a spinodal and critical point, and define a nonequilibrium chemical potential to interpret the ``binodal."  We provide a generalization of thermodynamic concepts like the free energy and temperature for nonequilibrium active systems.  Our theory agrees with existing simulation data both qualitatively and quantitatively and may provide a  framework for understanding and predicting the behavior of nonequilibrium active systems.
\end{abstract}

\maketitle

From bacteria swarms to fish schools to zebra herds, self-propulsion is a core feature of all ``active matter" systems.  By controlling and directing their own behavior self-propelled entities (usually, but not restricted to, living systems) can exhibit distinct phases with unusual dynamical properties\cite{Toner05}.  These exotic behaviors are made possible because active matter is an inherently nonequilibrium system that is not bound by conventional thermodynamic constraints.  A key challenge is to develop a framework for understanding the dynamic behavior and bulk properties of active matter.

While computer simulations have produced phase diagrams of active matter\cite{Stenhammar14,Speck14,Fily14,Wysocki14,Redner13}, many regions of phase space are difficult to explore because of the computational challenge of covering the parameter space.  In this paper we develop a new mechanical theory for predicting the phase behavior of active systems.  We also offer suggestions on how conventional thermodynamic concepts, such as chemical potential, free energy and temperature, can be extended to provide a `thermodynamics' of nonequilibrium active matter. Our analysis suggests that active systems are entropically driven  by a lower critical solution temperature (LCST) transition, where phase separation becomes possible with \textit{increasing} temperature.

\section{Mechanical Theory}
Recently, a swim pressure was introduced as a fundamental aspect of active  systems and as an aid to understand their large-scale collective behavior\cite{Takatori14,Takatori14b,Yang14}.  For a dilute system the ``ideal-gas" swim pressure is $\Pi^{swim} = n \zeta U_0^2 \tau_R / 6$, where $n$ is the number density of particles, $\zeta$ is the hydrodynamic drag factor, $U_0$ is the swim speed of an active particle, and $\tau_R$ is its reorientation time\cite{Takatori14}.

Dimensional analysis allows us to write the swim pressure as $\Pi^{swim}(k_s T_s, \phi, Pe_R) = n k_s T_s \widehat{\Pi}^{swim}(\phi, Pe_R)$, where $k_s T_s \equiv \zeta U_0^2 \tau_R / 6$ defines the swimmers' ``energy scale'' -- force ($\zeta U_0$) $\times$ distance ($U_0\tau_R$) -- and $\widehat{\Pi}^{swim}(\phi, Pe_R)$ is the nondimensional swim pressure that depends in general on the volume fraction $\phi = 4 \pi a^3 n / 3$ and the reorientation P\'eclet number $Pe_R = a/(U_0 \tau_R)$, the ratio of the swimmer size $a$ to its run length $U_0\tau_R$.

For large $Pe_R$ the swimmers reorient rapidly and take small swim steps, behaving as Brownian walkers\cite{Takatori14}.  Thus $\widehat{\Pi}^{swim}(\phi, Pe_R) = 1$ for all $\phi \lesssim \phi_0$ where $\phi_0$ is the volume fraction at close packing.  This system is analogous to passive Brownian particles, which exert the ``ideal-gas" Brownian osmotic pressure $\Pi^B = n k_B T$ regardless of the concentration of particles.

For small $Pe_R$ the swimmers have large run lengths compared to their size and $\widehat{\Pi}^{swim}$ decreases with $\phi$ because the particles hinder each others' movement.  In this limit experiments and computer simulations\cite{Theurkauff12,Palacci13,Redner13,Bialke13,Stenhammar13,Buttinoni13} have observed the self-assembly of active systems into dense and dilute phases resembling an equilibrium liquid-gas coexistence.

Extending the results of the nonlinear microrheology analysis\cite{Takatori14} the swim pressure at small $Pe_R$ in 3D takes the form $\widehat{\Pi}^{swim} = 1 - \phi - \phi^2$.  The inclusion of a three-body term $(-\phi^2)$ agrees with the swim pressure data for all $Pe_R \le 1$.  Unlike Brownian systems where repulsive interactions (e.g., excluded volume) increase the pressure, for active matter interactions decrease the run length and therefore the swim pressure.  The decrease in $\Pi^{swim}$ is the principle destabilizing term that facilitates a phase transition in active systems.  

At finite concentrations, interparticle interactions between the swimmers give rise to an interparticle (or collisional) pressure $\Pi^P(k_s T_s, \phi, Pe_R) = n k_s T_s \widehat{\Pi}^P(\phi, Pe_R)$, where $\widehat{\Pi}^P(\phi, Pe_R)$ is the nondimensional interparticle pressure.  For repulsive  interactions $\Pi^P$ increases monotonically with $\phi$ and helps stabilize the system.  The phase behavior of active systems is determined by a competition between a destabilizing $\Pi^{swim}$ versus a stabilizing $\Pi^P$, a balance controlled by the parameter $Pe_R$.

For large $Pe_R$ the swimmers behave as Brownian particles and $\widehat{\Pi}^P(\phi, Pe_R) = \widehat{\Pi}^{HS}(\phi)$, where $\widehat{\Pi}^{HS}(\phi) = 4 \phi g(2;\phi)$ is the interparticle pressure of hard-sphere Brownian particles\cite{Brady93} and $g(2;\phi)$ is the pair-distribution function at contact.  The detailed interactions between the particles are not important\cite{Brady93}---a hard-sphere molecular fluid's interparticle pressure has the same form -- the same volume fraction dependence -- as that of a Brownian system despite differences in the source of the collisions.  A system of active swimmers also exhibits the same form of the interparticle pressure. Indeed, for large $Pe_R$ the run length $U_0 \tau_R$ sets the scale of the force moment and $\Pi^P \sim n^2 \zeta U_0 a^3 (U_0\tau_R) \sim n k_s T_s \phi$, analogous to the passive hard-sphere Brownian collisional pressure $\sim n k_B T \phi$.

For small $Pe_R$, $\Pi^P \sim n^2 \zeta U_0 a^4 \sim n k_s T_s Pe_R \phi$ since a swimmer is displaced by its size $a$ upon collision, not the run length $U_0 \tau_R$.  The interparticle pressure for small $Pe_R$ in 3D is thus $\widehat{\Pi}^P = 3 \phi Pe_R g(2;\phi)$.  

For both small and large $Pe_R$,   the pair-distribution function at contact has the form $g(2;\phi) = \left( 1 - \phi/\phi_0 \right)^{-\beta}$, and $\phi_0$ and $\beta$ are parameters obtained from the interparticle pressure of hard-sphere molecular fluids and/or passive Brownian particles.  Simulations verify that the parameters $\phi_0 = 0.65$ and $\beta = 1$ agree independently with the collisional pressures for hard-sphere active swimmers, passive Brownian particles, and molecular fluids. 

\begin{figure}[t!]
	\begin{center}
		\includegraphics[width=1.05\linewidth]{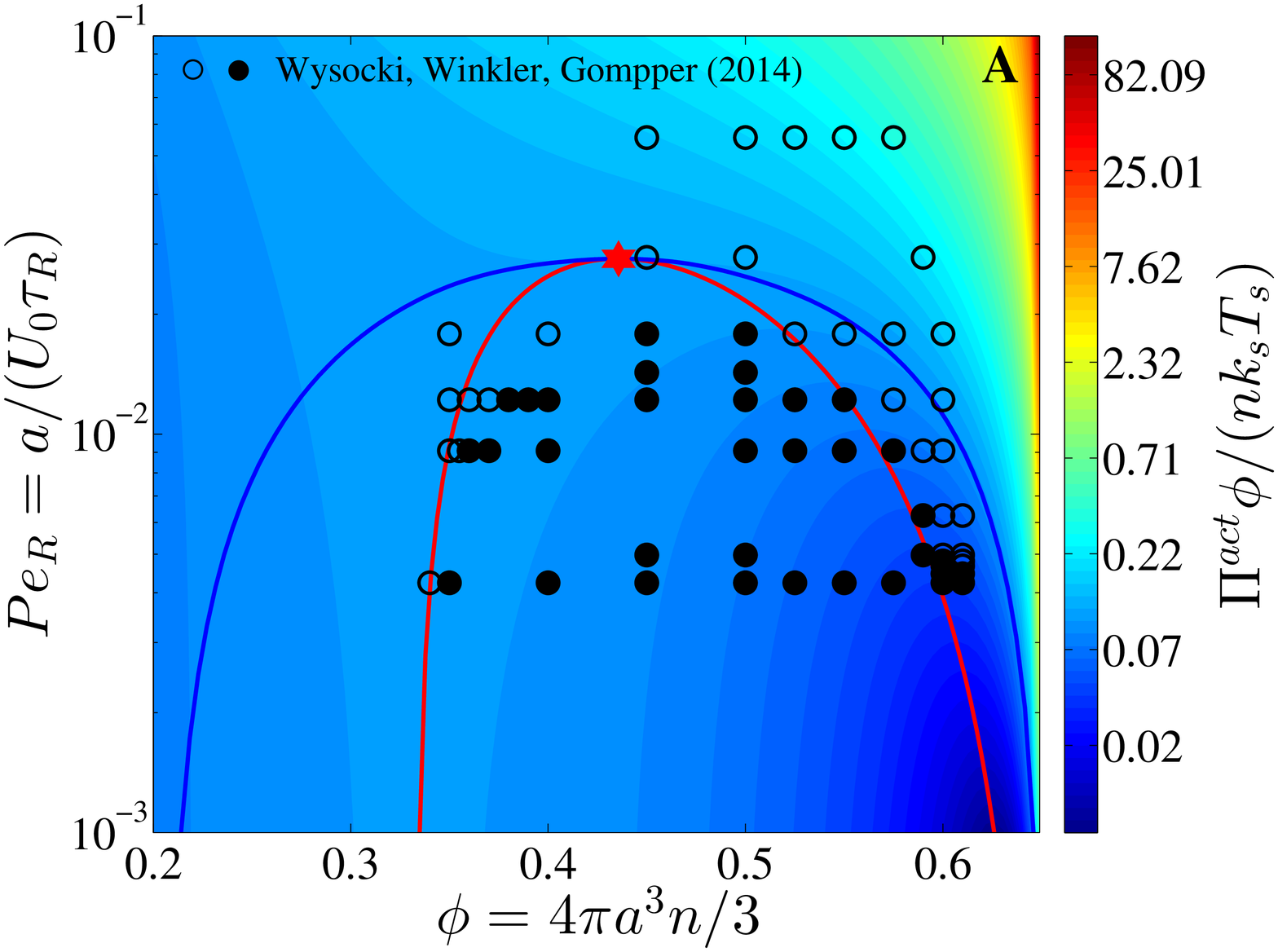} 
		\includegraphics[width=1.05\linewidth]{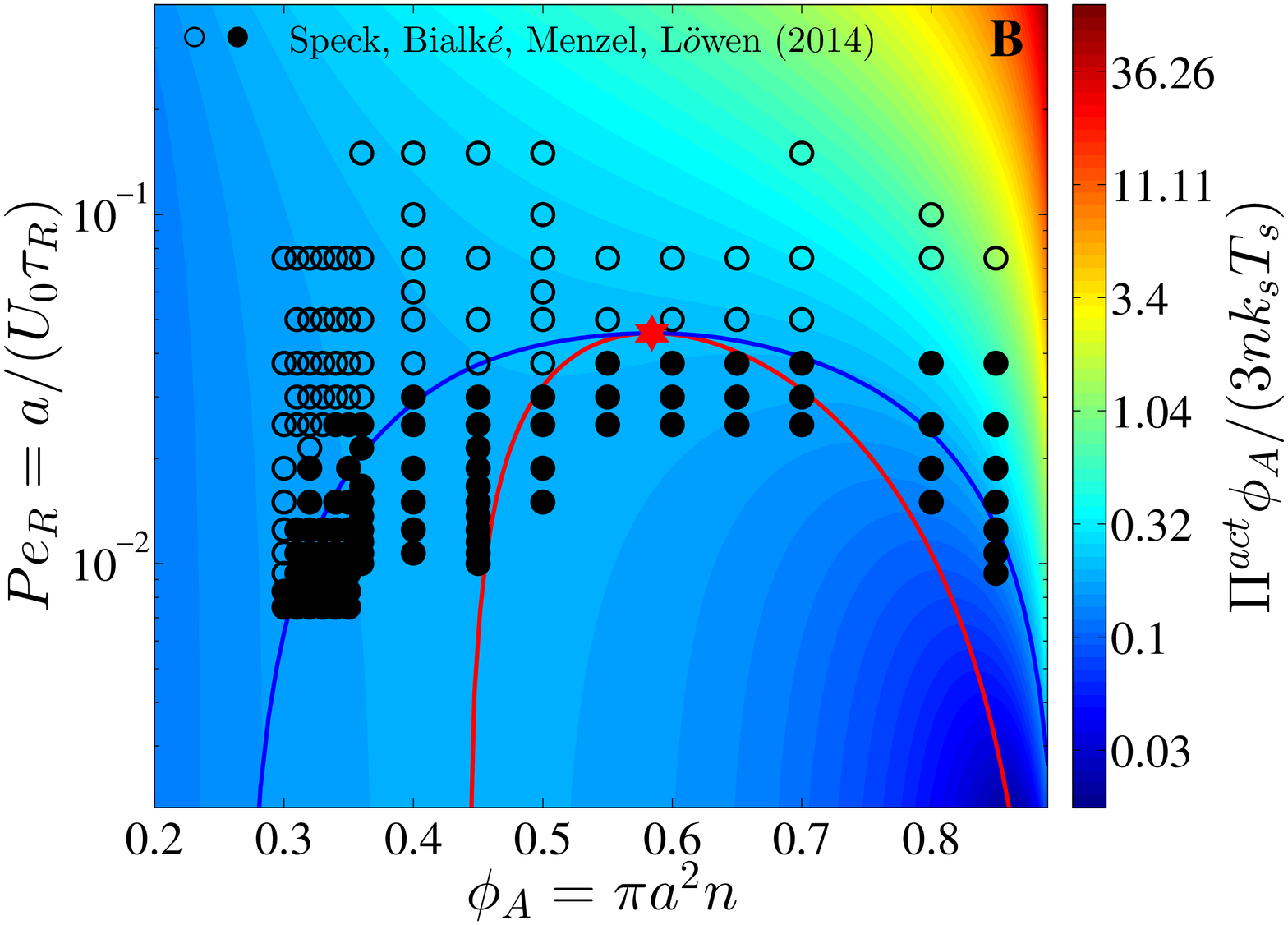} 
		\caption{Phase diagram in the $Pe_R-\phi$ plane in (\bd{A}) 3D and (\bd{B}) 2D.  The colorbar shows the active pressure scaled with the swim energy $k_s T_s = \zeta U_0^2 \tau_R / 6$, and the blue and red curves are the binodal and spinodal, respectively.  The critical point is shown with a red star.  The open and filled symbols are simulation data with a homogeneous and phased-separated state, respectively.} \label{Fig:1}
	\end{center}
\end{figure}

The active pressure is the sum of the swim and interparticle pressures\footnote{Pressure $p_f$ of the incompressible solvent is arbitrary.}, which  
for small $Pe_R$ is
\begin{equation} \label{eq:activepressure}
\Pi^{act} = nk_sT_s\left( 1 - \phi - \phi^2 + 3 \phi Pe_R(1- \phi/\phi_0)^{-1}\right),
\end{equation}
and which we can use to analyze the phase separation in active matter.  We focus on non-Brownian swimmers since the effect of translational Brownian diffusivity is small in phase-separating systems.  Figure \ref{Fig:1} compares the phase diagram in the $Pe_R-\phi$ plane obtained from this model to the simulation data of other studies.

The spinodal defines the regions of stability and is determined by setting $\partial \Pi^{act}/\partial \phi = 0$.  This is given by the red curve in Fig \ref{Fig:1} that passes through the extrema of each constant-pressure isocontour (``isobar").  No notion of free energy is needed to obtain the spinodal---it is a purely mechanical quantity.

At the critical point $\partial \Pi^{act}/\partial \phi = \partial^2 \Pi^{act}/\partial \phi^2 = 0$.  In 3D we find the critical volume fraction $\phi^c \approx 0.44$, active pressure $\Pi^{act,c} \phi^c/(n k_s T_s) \approx 0.21$, and P\'eclet number $Pe_R^c \approx  0.028$, values consistent with our BD simulations.  Like the spinodal, the critical point is identified using only mechanical arguments.

The blue curve in Fig \ref{Fig:1} delineates the ``binodal" or coexistence regions, which we define as the equality of the chemical potential in the dilute and dense phases.  Although the thermodynamic chemical potential is defined only for equilibrium systems, one can define a proper nonequilibrium chemical potential for active systems using standard macroscopic mechanical balances\cite{Takatori14}: $n (\partial \mu^{act}/\partial n) = (1-\phi) \partial \Pi^{act}/\partial n$.

This definition agrees with the true thermodynamic chemical potential for molecular or colloidal solutes in solution\cite{Doi13}.  There are no approximations other than incompressibility of the solvent. Stress-induced diffusion, which this relationship implies, has been used in the context of particle migration of non-Brownian particles in pressure-driven flow\cite{Nott94}.  We thus interpret $\mu^{act}$ as a natural definition and extension of the chemical potential for nonequilibrium systems, and use it to compute and define a ``binodal."  

For small $Pe_R$ we obtain
\begin{multline} \label{eq:1}
	\mu^{act}(k_s T_s, \phi, Pe_R) = \mu^\theta(k_s T_s, Pe_R) + k_s T_s \log \phi \ + \\ k_s T_s\log \Gamma(\phi,Pe_R),
\end{multline}
where $\mu^\theta(k_s T_s, Pe_R)$ is the reference state whose form is not needed, and $\Gamma(\phi,Pe_R)$ is a nonlinear but analytic expression\footnote{\begin{multline*} \Gamma(\phi,Pe_R) = \left( 1 - \phi/\phi_0 \right)^{-3 \phi_0 Pe_R} \exp\left[ \phi^3 - \phi^2/2 \  + \right.\ \\ \left.\
3 Pe_R \phi_0(1-\phi_0)/(1 - \phi/\phi_0) - 3 \phi (1 - \phi_0 Pe_R) \right] \end{multline*}}.  The second term on the right-hand side represents the entropic, ``ideal-gas" contribution to the chemical potential.  The third term is the nonideal term that is the analog of enthalpic attraction between the active swimmers, and is represented by the quantity $\Gamma(\phi,Pe_R)$ that resembles the fugacity coefficient in classical thermodynamics.  Equation \ref{eq:1} is similar to that proposed by Cates and coworkers\cite{Tailleur08,Stenhammar14} who argued that $\mu(n) = \log n + \log \upsilon(n)$ where $\upsilon(n)$ is a density-dependent swimmer velocity.  Although an analytical expression for $\upsilon(n)$ has been proposed for dilute concentrations\cite{Stenhammar13,Bialke13}, our theory gives the nonideal contribution $\Gamma(\phi,Pe_R)$ in the entire range of $\phi$ and $Pe_R$.

\begin{figure}[t!]
	\begin{center}
		\includegraphics[width=0.47\textwidth]{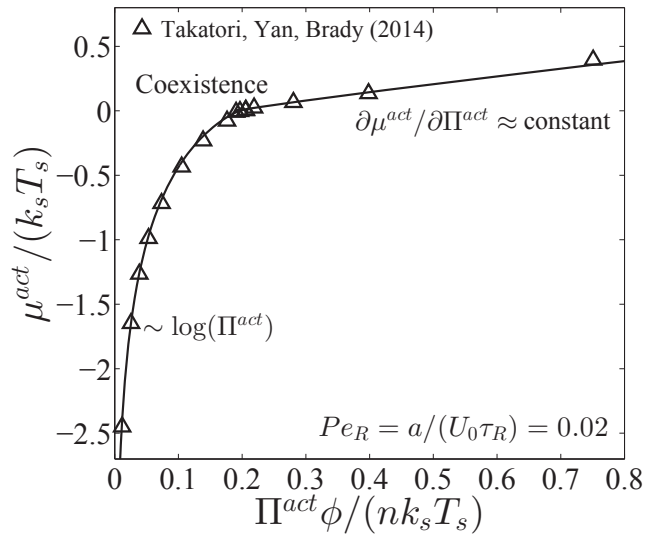} 
		\caption{Nonequilibrium chemical potential as a function of $\Pi^{act}$ for $Pe_R = a/(U_0 \tau_R) = \zeta U_0 a / (6 k_s T_s) = 0.02$, where $k_s T_s = \zeta U_0^2 \tau_R / 6$ is the swimmers' energy scale.  The symbols are BD simulations\cite{Takatori14} and the curve is the model, Eq \ref{eq:1}.} \label{Fig:2}
	\end{center}
\end{figure}

The chemical potential from BD simulations and the model is shown in Fig \ref{Fig:2} for $Pe_R = 0.02$.  It increases logarithmically at low $\Pi^{act}$ and the slope changes dramatically at the coexistence point ($\Pi^{act} \phi / (n k_s T_s) \approx 0.2$).  At this value of $\Pi^{act}$ and $Pe_R$ the chemical potentials are equal in the dilute and dense phases.  The data in the flat van der Waals region of the $\Pi^{act}-\phi$ phase diagram (see $\phi \approx 0.25 - 0.6$ in 
Fig 2 of \cite{Takatori14}) collapse onto the single coexistence point.

We can now define a ``binodal" in Fig \ref{Fig:1} through the equality of the chemical potential in both phases.  Our theory predicts that active systems prepared outside the binodal (blue curve) are stable in the homogeneous configuration and do not phase separate.  The regions between the spinodal and binodal are metastable and a homogeneous system does not spontaneously phase separate via spinodal decomposition but can undergo a nucleation process.  Nucleation times can be large and difficult to reach computationally, so artificial seeding may be required to induce phase separation\cite{Redner13}.

As shown in Fig \ref{Fig:1}A in 3D the transition from the homogeneous (open symbols) to phase-separated (filled symbols) systems in the simulations of Wysocki et al\cite{Wysocki14} agree well with the spinodal of our  model.

In 2D nucleation seeds form more easily, allowing nucleation processes to be more accessible in a simulation prepared near the binodal.  These observations are corroborated by Fig \ref{Fig:1}B where we take the swim and interparticle pressures in 2D as $\Pi^{swim}/(n \zeta U_0^2 \tau_R/2) = 1 - \phi_A - 0.2 \phi_A^2$ and $\Pi^P/(n \zeta U_0^2 \tau_R / 2) = (4/\pi) \phi_A Pe_R g(2;\phi_A)$, respectively, where $\phi_A = n \pi a^2$ is the area fraction of active swimmers and $g(2;\phi_A) = \left( 1 - \phi_A/\phi_0 \right)^{-\beta} $ with $\phi_0 = 0.9$ and $\beta = 1$.  The 2D simulation of Speck et al\cite{Speck14} show that the transition from the homogeneous (open symbols) to phase-separated (filled symbols) states occur near the binodal (blue curve).  

Our active pressure model agrees qualitatively and even quantitatively with the phase diagrams in Fig \ref{Fig:1}, as well as with those of other studies\cite{Stenhammar14,Fily14,Redner13}.  It should be appreciated that the predictions of our theory have no adjustable parameters.

\section{`Thermodynamic' Quantities}
The results presented thus far come from purely micromechanical arguments with no appeal to thermodynamics.  We now turn our attention towards thermodynamic properties like the free energy and temperature, which, although well-defined for an equilibrium system, have been elusive for nonequilibrium systems.  

Upon carefully imposing incompressibility of the solvent, one can relate the nonequilibrium Helmholtz FE to the mechanical pressure as $\Pi^{act}(k_s T_s, \phi, Pe_R) = \phi^2 \left[ \partial/\partial \phi \left((F^{act}/V)/\phi\right) \right]$, where $V$ is the volume of the system\cite{Doi13}.  There are again no approximations; it can be considered as the definition of the free energy for nonequilibrium active systems.  Substituting the active pressure model for small $Pe_R$ in 3D, we obtain 
\begin{multline} \label{eq:2}
	F^{act}/(N k_s T_s) = \log \phi - \phi(\phi + 2)/2  \ -\\ 3 Pe_R \phi_0 \log\left(1 - \phi/\phi_0\right)  +  F^\theta(k_s T_s, Pe_R), 
\end{multline}
where $N$ is the number of active swimmers and $F^\theta(k_s T_s, Pe_R)$ is the reference Helmholtz FE.  The first term on the right can be interpreted as the ideal entropic contribution, and the rest represent the nonideal ``enthalpic" attractions between the active swimmers.  For large $Pe_R$, the Helmholtz FE has no dependence on $Pe_R$: $F^{act}/(N k_s T_s) = \log \phi + 4 \int_0^\phi g(2;s) ds + F^\theta(k_s T_s, Pe_R)$.  The Helmholtz FE has a form in agreement with Cates and coworkers\cite{Tailleur08,Stenhammar14} who expressed the FE density as $f = n(\log n - 1) + \int_0^n \log \upsilon(s) ds$.

\begin{figure}[t!]
	\begin{center}
		\includegraphics[width=0.95\linewidth]{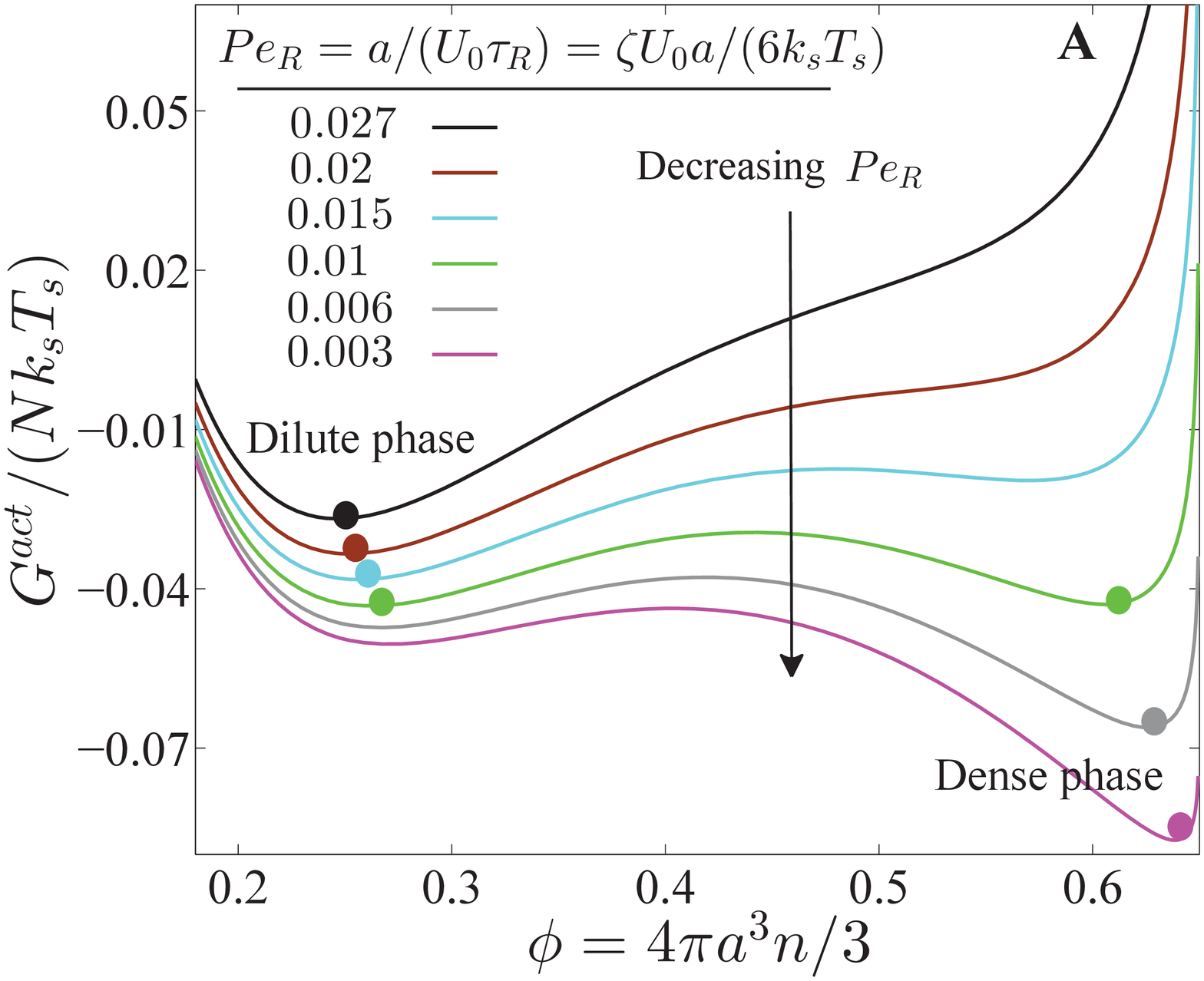} 
		\hspace*{\fill}
		\includegraphics[width=0.97\linewidth]{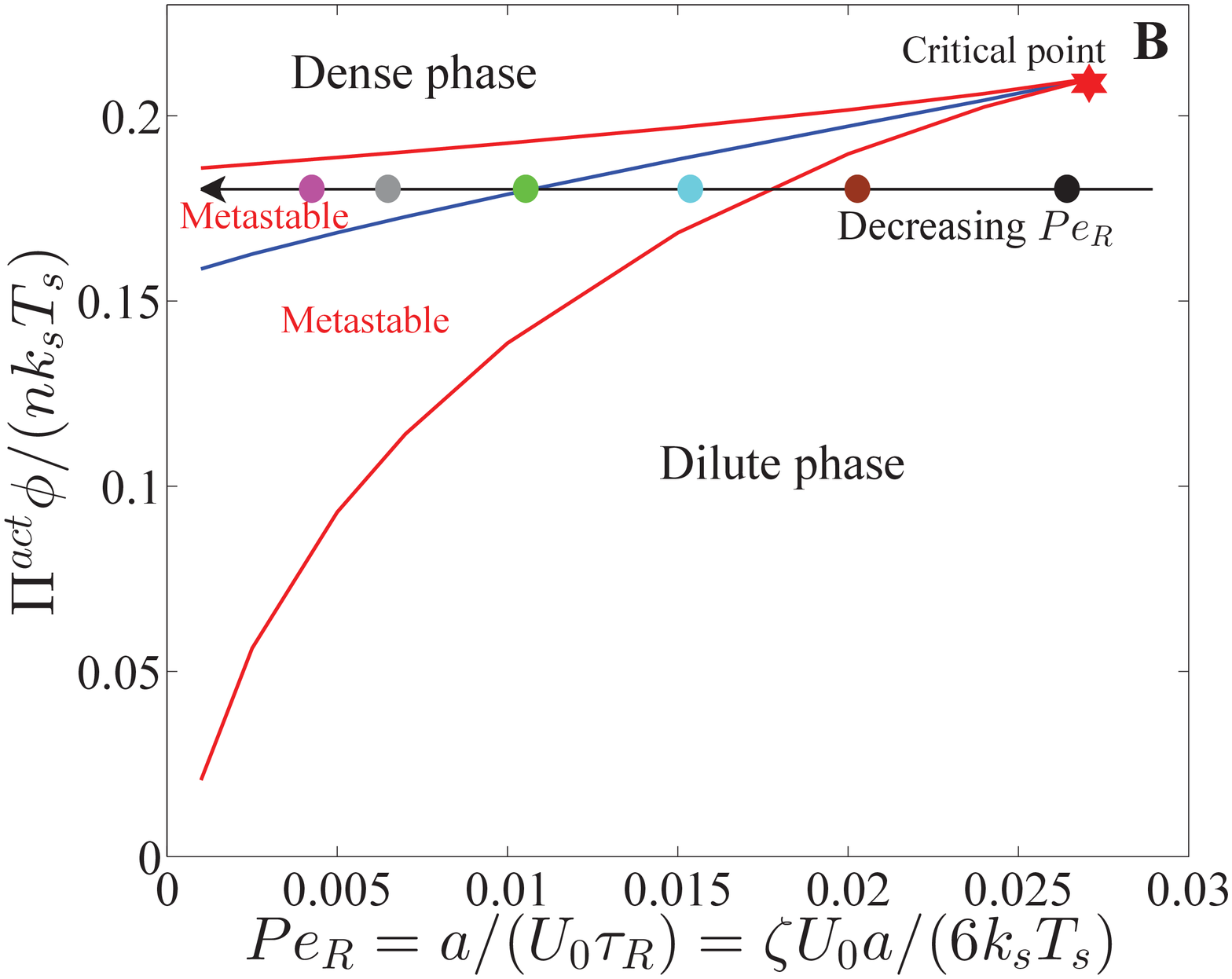} 
		\caption{(\bd{A}) Gibbs free energy (FE) as a function of $\phi$ for fixed values of $Pe_R$ and $\Pi^{act}\phi/(n k_s T_s) = 0.18$, where $k_s T_s = \zeta U_0^2 \tau_R / 6$. (\bd{B}) $\Pi^{act}-Pe_R$ phase diagram.  The red and blue curves are the spinodal and binodal, respectively.  The black arrow points towards decreasing $Pe_R$ at fixed $\Pi^{act}$.  The filled color circles in both graphs denote the stable states. } \label{Fig:3}
	\end{center}
\end{figure}

Given a chemical potential we can further define the Gibbs FE as $\mu^{act} = \left(\partial G^{act}/\partial N\right)_{N_f, \Pi^{act}, T_s, Pe_R}$, where $N_f$ is the number of solvent molecules\cite{Doi13}.  Alternatively we can compute the Gibbs FE from the Helmholtz FE\cite{Doi13}: $G^{act}/(N k_s T_s) = F^{act}/(N k_s T_s) + \Pi^{act}/(n k_s T_s)$.  Figure \ref{Fig:3}A shows the Gibbs FE as a function of $\phi$ for different values of $Pe_R$ and fixed $\Pi^{act}\phi/(n k_s T_s) = 0.18$.  This graph may be best understood alongside the $\Pi^{act} - Pe_R$ phase diagram in Fig \ref{Fig:3}B.  As $Pe_R$ decreases from a stable, dilute ``ideal gas" phase to $Pe_R = 0.015$ with a fixed $\Pi^{act} \phi / (n k_s T_s) = 0.18$, $G^{act}$ has a local minimum at $\phi \approx 0.6$ corresponding to the metastable dense phase (i.e., ``superheated liquid") and a global minimum at $\phi \approx 0.25$ corresponding to the stable dilute phase.  At $Pe_R = 0.01$ the two minima of $G^{act}$ are equal corresponding to the coexistence of the dilute and dense phases.

By writing the ``ideal-gas" swim pressure as $\Pi^{swim} = n \zeta U_0^2 \tau_R / 6 = n k_s T_s$, we can identify a swimmer's energy scale as $k_s T_s = \zeta U_0^2 \tau_R / 6$.  The reorientation P\'eclet number can be written as $Pe_R = a/(U_0 \tau_R) = \zeta U_0 a / (6 k_s T_s)$, which is interpreted as a ratio of the interactive energy of the swimmer -- the energy required to dispalce the swimmer its size -- 
to the swim energy scale $k_s T_s$.  Analogous to the Stokes-Einstein-Sutherland relation, one can interpret the swim diffusivity as $D^{swim} = k_s T_s / \zeta$, which also gives $Pe_R = U_0 a / D^{swim} \sim \zeta U_0 a / (k_s T_s)$.

From Fig \ref{Fig:1} phase separation occurs for small $Pe_R = \zeta U_0 a / (6 k_s T_s)$, or high $T_s$.  This is opposite to what is typically observed in a classical thermodynamic system, where phase separation is driven by attractive enthalpic interactions and becomes possible at low temperatures.  Phase separation with \textit{increasing temperature} is uncommon but has been observed for systems driven by the lower critical solution temperature (LCST) transition\cite{Griffiths70,Freeman60} where phase transition is dominated by entropy.  As $Pe_R$ decreases ($T_s$ increases) and the run length of the swimmer increases, the particle effectively becomes larger in size and thus has less space available for entropic mixing. 

Unlike a molecular fluid particle that can transmit its kinetic activity to another particle upon collisions, a swimmer cannot impart its intrinsic activity to another swimmer.  The motion of an inactive bath particle (i.e., neither active nor Brownian) in a dilute suspension of active swimmers is characterized by the diffusivity $D^{bath} \sim \phi U_0 a$, where $\phi$ is the volume fraction of the swimmers.  The ratio of the bath to swimmer diffusivity is $D^{bath}/D^{swim} \sim \phi U_0 a/(U_0^2 \tau_R) = \phi Pe_R$, suggesting that $Pe_R$ is the quantity that gets shared between the swimmers and not the swim energy $k_s T_s$.  

The entropy of active matter can be defined as $S^{act} = - \left( \partial G^{act}/\partial T_s \right)_{\zeta U_0 a, \Pi^{act}} =  - \left( \partial F^{act}/\partial T_s \right)_{\zeta U_0 a, \phi}$.  Ignoring the reference states, for large $Pe_R$ the entropy has the same form as that for a passive Brownian system: $S^{act}/ (N k_s) = -\log \phi - 4 \int_0^\phi g(2;s) ds$.  For small $Pe_R$ the entropy comes solely from the swim pressure: $S^{act}/(N k_s) = - \log \phi + \phi (\phi+2)/2$.  The entropy decreases with $\phi$ since the swimmers have less space available for entropic mixing.

The heat capacity can be obtained from $C_V = -T_s (\partial^2 F^{act}/\partial T_s^2)_{\phi, \zeta U_0 a}$.  Aside from the reference state, substitution of the FE into this equation gives $C_V = 0$ for all $\phi$ at both small and large $Pe_R$.  A possible explanation is that active matter has no true notion of the internal energy---since the swimmers cannot exchange their swim energy $k_s T_s$, there is no heat exchange between ``hot" (high activity) and ``cold" (low activity) active systems.  There is no ``first law" of thermodynamics for active matter systems.\footnote{The actual chemical energy consumed in propelling the swimmer is dissipated into the thermal bath of the solvent.  The behavior an active system depends on the activity $\zeta U_0$, not the actual energy consumed.}

In experimental systems the swimmers may achieve motion by consuming and converting chemical fuel.  If we allow for a density-dependent intrinsic swim speed ${U}_0(\phi)$ and reorientation time $\tau_R(\phi)$, our definition of the nonequilibrium chemical potential becomes
\begin{equation} \label{eq:3}
	n \frac{\partial \mu^{act}}{\partial n} = (1-\phi) \left[ \frac{\partial \Pi^{act}}{\partial n} - \Pi^{swim} \left( \frac{\partial \log (U_0 \tau_R)}{\partial n} \right) \right].
\end{equation}
If we had a nonzero average external force $\langle \vec{F}^{ext} \rangle$ derivable from a potential $V^{ext}$, $\mu^{ext} = V^{ext}$ must be added as a separate contribution to the total chemical potential of the swimmers.  Since $\Pi^{act}$ was determined for a homogeneous system, Eq \ref{eq:activepressure} still applies, but now $k_sT_s$ is also a function of $\phi$.

In active systems the relevant length scale is the swimmers' run length $U_0\tau_R$ and this must to be small compared to the apparatus size in an experiment for the continuum approach to hold.  In practice experiments may have non-continuum and non-local effects that may need to be considered when comparing experimental results with the thermodynamic model presented here.

Much work remains to explore the implications of our `thermodynamics' of active matter and to see if it might apply to other far from equilibrium systems.

\bigskip

\begin{acknowledgments}
SCT is supported by a Gates Millennium Scholars fellowship and a National Science Foundation  Graduate Research Fellowship (No. DGE-1144469).  This work is also supported by NSF Grant CBET 1437570.
\end{acknowledgments}

\bibliography{Diffusion}

\end{document}